# Improved S-AF and S-DF Relaying Schemes Using Machine Learning Based Power Allocation Over Cascaded Rayleigh Fading Channels

Yahia Alghorani, *Member, IEEE*, Ahmed Salim Chekkouri, *Member, IEEE*, Djabir Abdeldjalil Chekired, *Student Member, IEEE*, and Samuel Pierre, *Senior Member, IEEE*

*Abstract*—We investigate the performance of a dual-hop inter-vehicular communications (IVC) system with relay selection strategy. We assume a generalized fading channel model, known as cascaded Rayleigh (also called $n$*Rayleigh), which involves the product of $n$ independent Rayleigh random variables. This channel model provides a realistic description of IVC, in contrast to the conventional Rayleigh fading assumption, which is more suitable for cellular networks. Unlike existing works, which mainly consider double-Rayleigh fading channels (i.e, $n = 2$); our system model considers the general cascading order $n$, for which we derive an approximate analytic solution for the outage probability under the considered scenario. Also, in this study we propose a machine learning-based power allocation scheme to improve the link reliability in IVC. The analytical and simulation results show that both selective decode-and-forward (S-DF) and amplify-and-forward (S-AF) relaying schemes have the same diversity order in the high signal-to-noise ratio regime. In addition, our results indicate that machine learning algorithms can play a central role in selecting the best relay and allocation of transmission power.

*Index Terms*—$n$*Rayleigh distributions, machine learning.

## I. INTRODUCTION

**T**HE realization of inter-vehicular communications (IVC) is very challenging in practice and existing solutions, for example, from cellular and ad-hoc networks may not be applicable, which is mainly due to the dynamic nature of wireless links and the involved mobility patterns. In general, mobile-to-mobile (M2M) fading channels often exhibit greater dynamics and more severe fading than fixed-to-mobile (F2M) cellular radio channels, which are mostly limited to the classical Rayleigh or Nakagami-$m$ distribution (i.e., $n = 1$), where

Manuscript received November 19, 2018; revised June 17, 2019, October 17, 2019, March 14, 2020, and June 6, 2020; accepted June 10, 2020. This work was supported by the Science Foundation Ireland (SFI) under Grant 13/CDA/2199. The Associate Editor for this article was G. Mao. *(Corresponding author: Yahia Alghorani.)*

Yahia Alghorani is with the Department of Electrical Engineering, Lakehead University, Thunder Bay, ON P7B 5E1, Canada (e-mail: yahia.alghorani@ieee.org).

Ahmed Salim Chekkouri is with Meduuicom, Ecole Polytechnique de Montreal, Montreal, QC H3C 3A7, Canada (e-mail: ahmed-salim.chekkouri@polymtl.ca).

Djabir Abdeldjalil Chekired is with the ICD/ERA, University of Technology of Troyes, 10300 Troyes, France (e-mail: djabir.chekired@utt.fr).

Samuel Pierre is with the Department of Computer Engineering, Ecole Polytechnique de Montreal, Montreal, QC H3C 3A7, Canada (e-mail: samuel.pierre@polymtl.ca).

Digital Object Identifier 10.1109/TITS.2020.3003820

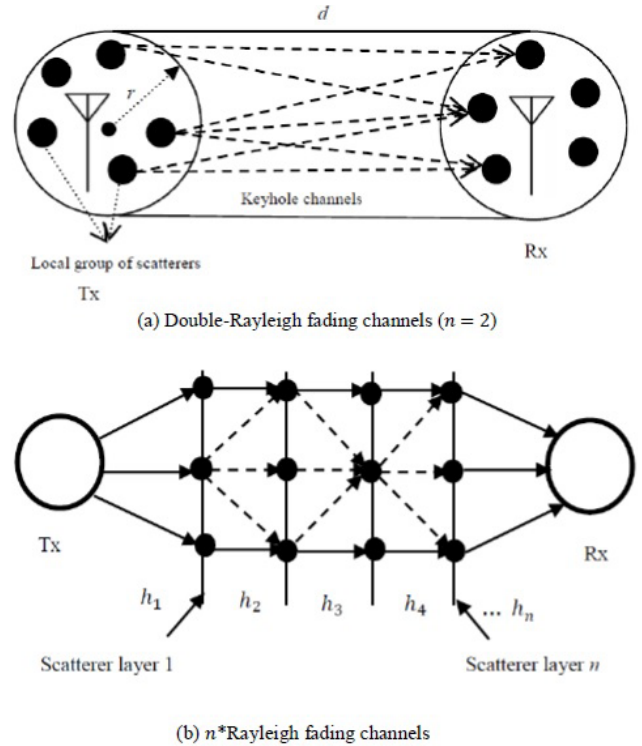

Fig. 1. Multiple-scattering model for M2M channels (where several factors can contribute to generate $n$*Rayleigh fading channels between two vehicles, e.g., when i) $d \gg r$, ii) there are $n$ AF relays between the transmitter and receiver (i.e., $n + 2$ local groups of scatterers), or iii) propagation paths exist).

the stationary base station has high elevation antennas and is relatively free from local scatterers [1]. Therefore, it is important to utilize a realistic channel model that characterizes the statistical properties of M2Mchannels such as $n$*Rayleigh distribution [2]. In IVC, both transmitter and receiver are in motion, and typically have the same antenna height, resulting in two or more small-scale fading processes generated by independent groups of local scatterers around the two mobile terminals [3] (see for example Fig.1, where multiple scattering is taken place between the transmitter and receiver, and all propagation paths travel through the same narrow pipe called keyhole channels). Such stochastic processes are widely encountered in dense urban and forest environments where local scattering objects such as buildings, vehicles, street corners, road signs, tunnels, hallways, bridges, trees and mountains, obstruct a direct radio wave path between two vehicles leading to nonline- of-sight (NLOS) propagation paths [4]. Depending on the type of an obstructing surface, the transmitted radio signal may undergo reflection, diffraction, or scattering, resulting in fast or slow fading, which in turn leads to deterioration of link reliability (e.g., high outage probability and low data rate), an increase in the number of connection drops, and a decrease in battery life. However, there are other forms for the keyholes in a realistic environment that arise in multi-hop amplify-and-forward (AF) relay networks [5], [6]. The AF relaying system basically works as a keyhole when it amplifies the received signal; in a sense that the amplitude of the received signal is a product of $n*$Rayleigh random variables (e.g., the double-Rayleigh signal amplitude in F2M scenarios [7]). A similar behaviour can also be found when two rings of local scatterers around a transmitter and a receiver are separated by a large distance and the radio wave passes through the keyhole channels [8]. The $n*$Rayleigh channel model is classified as a multi-scattering channel model, in which the "keyhole[1]" contributions are summed together to give a generalization of a single scattering (Rayleigh) model [3].

## II. Related Work

In [9] and [10], M2M channel statistics were discussed, such as the probability density function (PDF) and the cumulative distribution function (CDF) of the product of $n*$Rayleigh random variables. In [11], experimental results in different communication environments have shown that if several small-scale fading processes are multiplied together, the worse-than Rayleigh fading model is generated. In [12], multiple-input multiple-output (MIMO) antenna systems were investigated through $n*$Rayleigh fading channels. The study concluded that when the distance separation between the transmitter and the receiver is much greater than the ring radii around the two terminals, a double-Rayleigh model ($n = 2$) is considered instead of a single-Rayleigh model. For M2M channel modelling, [13] characterized M2M channels in the 5-GHz band through measurement-calibrated ray-tracing models (e.g., the path loss, shadow fading, and delay spread of the channel) and showed agreement with measured results in the literature for all these channel characteristics. The ray-tracing approach is generally computationally intensive and sacrifices accuracy if computational complexity is reduced [14]. In [15], channel statistics, such as the time-variant space correlation functions, and corresponding Doppler power spectral density, were studied for three-dimensional non-stationary geometric models for M2M communications. The study showed that the analytic results are consistent with measured data. Although the geometric models can be used to accurately simulate the M2M scattering channels, they are very complex and require numerous parameter selections for the specific environment of interest [16]. Another recent study has been implemented in MIMO systems and antenna selection via $n*$Rayleigh fading channels [17], where the IVC systems achieve good cost-performance trade-off when the number of RF chains

(e.g., digital-to-analog and analog-to-digital converters) is limited. [18] presented an information-theoretic analysis of a point-to-point MIMO link affected by Rayleigh fading and multiple scattering, under perfect channel state information (CSI) at the receiver. The study analysed the sum-rate performance when the zero-forcing and mean-squared error receivers are adopted and suggested that the linear receivers are not well-suited for multi-fold scattering. Several studies have reported that cooperative communications through multiple Rayleigh fading channels can improve the link reliability when traffic density is high [6], [19]. Relay selection has been studied extensively in the literature, see, e.g., [20]–[22] and the references therein. However, current results are limited to Rayleigh fading channel assumption (i.e., $n = 1$). A few studies have discussed cooperative communications with the relay selection strategy via double Rayleigh fading channels (i.e., $n = 2$); see, e.g., [23] and [24],where the maximum achievable diversity order is equivalent to the number of relays. In order to understand the full potential of cooperative diversity in IVC, an in-depth analysis of the system performance under a realistic channel model is required. To this end, we investigate the IVC systems with several relay selection strategies, such as the selective decode-and-forward (S-DF) and amplify-and-forward (S-AF) relaying with the $n*$Rayleigh distribution, which to the best of our knowledge, have not been studied before. Therefore, it is the aim of this work to fill this research gap. Specifically, our main contributions through this work can be summarized as follows:

- We introduce new approximate analytical expressions but accurate for the outage probability for both S-AF and S-DF relaying schemes over cascaded Rayleigh fading channels.
- We propose a machine learning-based power allocation scheme to optimize the transmit power between the source and the selected relay.
- We demonstrate that the S-DF and S-AF relay schemes have the same maximum diversity order ($d = mN/n$)the high signal-to-noise ratio (SNR) regime, which degrades by increasing the cascading order ($n$) and improves by increasing the number of relays $N$.

## III. System and Channel Model

Consider a dual-hop cooperative IVC network with multiple relays (as shown in Fig.2), where a source ($s$), relays $r_i$ ($i = 1, \ldots, N$) and a destination ($d$) are equipped with a single pair of transmit and receive antennas, and operate in half-duplex mode. For each time instant, only one vehicle acts as a source, while the other vehicles serve as relays that help forward the source's message to the destination. To simplify notation, in the sequel we use the subscript '1' to represent the source-relay link and the subscript '2' to represent the relay-destination link. Here, all underlying channels between $s \rightarrow r_i$ and $r_i \rightarrow d$ links are modeled by a product of $n$ independent complex Gaussian random variables, each of which is defined by $h_{i1} \triangleq \prod_{k=1}^{n_{i1}} h_{i1,k}$ and $h_{i2} \triangleq \prod_{k=1}^{n_{i2}} h_{i2,k}$ Hence, $|h_{i1}|$ and $|h_{i2}|$ follow an $n*$Rayleigh distribution. In this system model, we assume that all underlying channels are quasi-static (i.e., slow fading) which can be justified for IVC scenarios in rush-hour traffic (e.g., urban environments where the average speed is low). We further assume that the additive white Gaussian noise (AWGN) at the relays and destination have zero mean and variance ($N_o$). The instantaneous SNRs of the links $s \rightarrow r_i$ and $r_i \rightarrow d$ are given, respectively, by $\gamma_{i1} = |h_{i1}|^2 P/N_o$ and $\gamma_{i2} = |h_{i2}|^2 P/N_o$, where $P$ is the radio transmit power of the source signal, which we shall initially assume to be equal to that transmitted from the selected relay. The PDF of instantaneous SNRs is given by [5]

$$f_{\gamma_{ij}}(\gamma) = \frac{1}{\gamma} G_{0,n_{ij}}^{n_{ij},0} \left( \frac{\gamma}{\bar{\gamma}_{ij}} \middle| \begin{matrix} - \\ 1,\dots,1 \end{matrix} \right), \quad j = 1, 2 \tag{1}$$

where $G_{p,q}^{m,n}$ (.) is the Meijer-G function which is defined in [25 eq. (9.301)], $\bar{\gamma}_{ij} = \lambda_{ij} P/N_o$ and $\lambda_{ij} = \mathbf{E}\left(|h_{ij}|^2\right)$.

---

[1]Here we define the keyhole as a multiplier between two fading processes, resulting in a received signal amplitude that is a product of n Rayleigh random variables [5], e.g., double-Rayleigh fading [8].

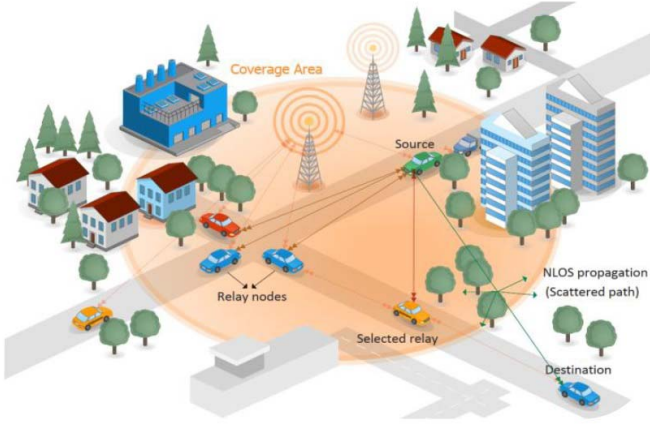

Fig. 2. Dual-hop cooperative IVC network in dense and high traffic scenarios, where the best relay is selected based on the channel propagation conditions (i.e., $n$*Rayleigh fading channels) over the source-relay-destination links.

### A. S-DF Relaying

In relay selection, two orthogonal time slots are utilized to perform the cooperative transmission. For example, in the S-DF relaying scheme, in the first time slot, the source transmits a message $x_s$ (where the source symbol is generated from a unit-energy complex constellation, such as phase-shift keying (PSK) and quadrature amplitude modulation (QAM) schemes) to a set of relay nodes and the destination. In this stage, the S-DF relaying policy is applied to choose the most reliable path over $s \rightarrow r_i \rightarrow d$ links. We define the decoding set ($D$) as the set of relays that decode the source symbol successfully; that occurs when the channel quality between the source and relay node is sufficiently good. Here, we assume that each relay can determine whether the source message is decoded correctly or not through a cyclic redundancy check (CRC). In the second time slot, only one relay from the decoding set, having the best link quality with the destination, will forward the estimate of the source symbol, denoted by $x_r$. Thus, the signal received by the relay node from the source is $y_{i1} = h_{i1}\sqrt{P}x_s + w_i$, and that received by the destination node from the selected relay is $y_2 = h_{i2}\sqrt{P}x_r + w_2$, where

$w_i$ and $w_2$ are the AWGN at the relay and destination nodes respectively.

For purpose of analysis, we generate a set of $N$ independent variates, each with CDF $F_\gamma(\gamma)$. Let the corresponding order statistics be denoted as $\gamma_N \geq \cdots \gamma_i \geq \cdots \geq \gamma_1$, where the relay selection process depends on the $i$-th order statistics $\gamma_{(i)}$ in samples of size $N$. Such a technique is useful in case that the best relay is connected by another source node or the SNR over $r_i \rightarrow d$ link suddenly deteriorates due to the impact of $n$*Rayleigh channels. In this case, the underlying protocol has to choose another relay instead to implement the transmission process. However, for any selected relay, the CDF of the $i$-th-order statistics $\gamma_{(i)}$ is given by [26]

$$F_{(i)}(\gamma) = \Pr\left(\gamma_{(i)} \leq \gamma\right)$$
$$= \sum_{k=i}^{N} \frac{1}{k!(N-k)!} \operatorname{per} \begin{pmatrix} F_1(\gamma) & 1 - F_1(\gamma) \\ \vdots & \vdots \\ F_N(\gamma) & 1 - F_N(\gamma) \end{pmatrix} \tag{2}$$

where per $(A)$ denotes the permanent of the $N \times N$ matrix $A$, which is defined in [26]. The matrix $A$ is obtained by taking $k$ copies of the vector $a_1$, and $N - k$ copies of the vector $a_2$ where $a_1$ and $a_2$ are the column vectors of $A$. In order to derive the CDF of the received SNR at the destination via the $s \rightarrow r_i \rightarrow d$ link $F_{\gamma_i}(\gamma)$, we invoke the technique described in [27], thus, the conditional PDF of the received SNR indicating that $r_i$ is idle when the instantaneous SNR of $s \rightarrow r_i$ link is below a threshold value ($\gamma_o = 2^{2}$ $^R - 1$, where $R$ is the target rate); is expressed as $f_{\gamma_i|r_i \text{ is off}}(\gamma) = \delta(\gamma)$, where $\delta(\gamma)$ is the Dirac delta function. Hence, the probability the $i$-th relay will not be in the decoding set $\mathcal{D}$ can be found as $\Pr(\gamma_{i1} \leq \gamma_o) = F_{\gamma_{i1}}(\gamma_o)$. On the other hand, the probability that $i$-th relay is in the decoding set is $(1 - \Pr(\gamma_{i1} \leq \gamma_0))$, and the conditional PDF given $r_i$ is active is $f_{\gamma_i|r_i \text{ is on}}(\gamma) = f_{\gamma_{i2}}(\gamma)$. Therefore, the CDF of the instantaneous end-to-end SNR via the $i$-th link is expressed as

$$F_{\gamma_i}(\gamma) = F_{\gamma_{i1}}(\gamma_o) + \left[1 - F_{\gamma_{i1}}(\gamma_o)\right] F_{\gamma_{i2}}(\gamma) \tag{3}$$

By replacing (3) in (2), we can calculate the CDF of the $i$-th order statistics $\gamma_{(i)}$. It is worth mentioning that (3) is complex due to the existence of Meijer-G function which requires high computational complexity, therefore, we adopt an approximate solution for the PDF in (1) to be expressed as [10]

$$f_{\gamma_{ij}}(\gamma) = \frac{\beta_{ij}^{m_{ij}}}{n_{ij}\,\Gamma\left(m_{ij}\right)} \gamma^{\alpha_{ij}-1} e^{-\beta_{ij}\gamma^{1/n_{ij}}} \tag{4}$$

where $\alpha_{ij} = m_{ij}/n_{ij}$, $\beta_{ij} = m_{ij}/\Omega_{ij} p_{ij}^{1/n_{ij}}$, and $n$*Rayleigh fading severity parameters are set as $m_{ij} = 0.6102\ n_{ij} + 0.4263$ and $\Omega_{ij} = 0.8808\ n_{ij} - 0.9661 + 1.12$. Now based on (4), we will be able to analyze the performance of the underlying schemes; especially in terms of the diversity order and power control as we see in the following sections. In order to obtain the PDF for the SNR in (4), we used the change of variable $f_\gamma(\gamma) = f_h(\sqrt{2^n\sigma^2\gamma/\bar{\gamma}})/2\sqrt{\gamma\ \bar{\gamma}/2^n\sigma^2}$, given in [28] with replacing the factor $2\sigma^2$ by $2^n\sigma^2$, where $\sigma^2 = \prod_{k=1}^{n}\sigma_k^2$ is the standard deviation of the original complex Gaussian signal prior to envelop detection. Using the facts given in [25, eq. (3.381.1) and eq. (8.356.3)] the approximate CDF for (4) is found as

$$F_{\gamma_{ij}}(\gamma) = 1 - \frac{\Gamma\left(m_{ij}, \beta_{ij}\gamma^{\frac{1}{n_{ij}}}\right)}{\Gamma\left(m_{ij}\right)} \tag{5}$$

where $\Gamma(\alpha, x) = \int_x^\infty e^{-t}t^{\alpha-1}dt$ represents the upper incomplete gamma function [25]. Similarly, using (5), (3), and (2), the CDF of $\gamma_{(i)}$ is obtained. In addition, we can derive a closedform expression for the CDF of the largest of $N$ random variables ($\gamma_{(N)}$) distributed according to (4), to be expressed as $\Pr\left(\gamma_{(N)} \leq \gamma\right) = \prod_{i=1}^N \left[F_{\gamma_i}(\gamma)\right]$, which is also obtained by using (3) and (5), as

$$F_{\gamma_{(N)}}(\gamma) = \prod_{i=1}^N \left[ 1 - \frac{\Gamma\left(m_{i1}, \beta_{i1}\gamma_o^{\frac{1}{n_{i1}}}\right)\Gamma\left(m_{i2}, \beta_{i2}\gamma^{\frac{1}{n_{i2}}}\right)}{\Gamma\left(m_{i1}\right)\Gamma\left(m_{i2}\right)} \right] \tag{6}$$

### B. S-AF Relaying

For the S-AF relaying scheme, particularly, in the second time slot, only the selected relay with the maximum effective SNR is chosen to forward the amplified received signal $x_r$ to the destination with a channel gain $G = \sqrt{1/\left(P|h_{i1}|^2 + N_o\right)}$. In this case, the signal received by the destination is expressed as $y_2 = h_{i2}\sqrt{P}x_r + w_2$, where $x_r = G y_{i1}$. Thus, the effective end-to-end SNR for the selected relay, can be upper-bounded as [29]

$$\gamma_{(H,N)} \leq \max_i \frac{\gamma_{i1}^{\frac{1}{n_{i1}}} \gamma_{i2}^{\frac{1}{n_{i2}}}}{\gamma_{i1}^{1/n_{i1}} + \gamma_{i2}^{1/n_{i2}}} \tag{7}$$

where $n_{i1} = n_{i2} = n_i$. Since the AF relaying schemes consider the end-to-end SNR for each relay ($\gamma_i$) compared to the DF relaying, we can presume that both links of $s \to r_i$ and $r_i \to d$ have the same cascading order $n$ to simplify the analysis. Having said that, we are interested in knowing the total value of $n$ generated between the source and the destination. Using the definition of the harmonic mean of two random variables [30], given as $\mu_H(X_1, X_2) = 2\,X_1X_2/(X_1 + X_2)$, (7) can be rewritten as $\gamma_{(H,N)} = \max_i \left\{\gamma_{H,i}\right\}$, where $\gamma_{H,i} = \mu_H\left(\gamma_{i1}^{1/n_i}, \gamma_{i2}^{1/n_i}\right)/2$. However, it is worthwhile to note that the derivation of the outage probability for the S-AF scheme which is based on (1) and (7) does not lend itself to a closed-form solution. Hence, to simplify the analysis, we use the approximate PDF given in (4).

## IV. DERIVATION OF PDF AND CDF FOR THE HARMONIC SNR

In order to find the PDF and CDF of the harmonic SNR $-\gamma_{H,i} = \gamma_{i1}^{1/n_i} \gamma_{i2}^{1/n_i}/\left(\gamma_{i1}^{1/n_i} + \gamma_{i2}^{1/n_i}\right)$ when the average links SNR ($\gamma_{i1}, \gamma_{i2}$) are i.i.d random variables, we introduce the following proposition:

*Proposition*: Suppose $Y_1$ and $Y_2$ are two i.i.d. gamma random variables, defined as $Y_1 = X_1^{\frac{1}{n}}$ and $Y_2 = X_2^{\frac{1}{n}}$

(where the RV $X_j = \gamma_j$ has an $n$*Rayleigh distribution, $j = 1, 2$ and $n \in \mathbb{N}^+$ ) with parameters $n\alpha > 0$ and $\beta > 0$ (i.e., $Y_j \sim G(n\alpha, \beta)$), the PDF and CDF of the harmonic mean of the two gamma RVs, $Y = \mu_H(Y_1, Y_2)$, can be expressed as

$$f_Y(y) = \frac{\sqrt{\pi}\beta}{2^{2(n\alpha-1)}\Gamma^2(n\alpha)} G_{1,2}^{2,0}\left(2\beta y \Big|\begin{array}{c} n\alpha - \frac{1}{2} \\ n\alpha - 1, 2n\alpha - 1, -1 \end{array}\right) \tag{8}$$

and

$$F_Y(y) = \frac{\sqrt{\pi}\beta y}{2^{2(n\alpha-1)}\Gamma^2(n\alpha)} G_{2,3}^{2,1}\left(2\beta y \Big|\begin{array}{c} 0, n\alpha - \frac{1}{2} \\ n\alpha - 1, 2n\alpha - 1, -1 \end{array}\right) \tag{9}$$

respectively.

*Proof:* Since the approximate PDF of the $n$*Raleigh random variable, $X_j$, is given by (4), $f_{X_j}(x) = \frac{\beta^m}{n\Gamma(m)}x^{\alpha-1}e^{-\beta x^{\frac{1}{n}}}$, the PDF of the RV $Y_j = X_j^{\frac{1}{n}}$ can be found with the help of [31, Sec. 5.2] as $f_{Y_j}(y) = \frac{\beta^{n\alpha}}{\Gamma(n\alpha)}y^{n\alpha-1}e^{-\beta y}$, hence, the RV $Y_j$ follows a gamma distribution with parameters $(n\alpha, \beta)$. In order to calculate the PDF of the harmonic mean of $Y_1$ and $Y_2$, $Y = 2\,Y_1Y_2/Y_1 + Y_2$, we define the following two RVs as

$$U = 2Y_1Y_2$$
$$V = Y_1 + Y_2 \tag{10}$$

Now, taking the Jacobian transformation of (11), with the help of [32, eq. (07.34.21.0085.01) and eq. (07.34.21.0084.01)] and (9) can be proved.

Using the fact that $G_{1,2}^{2,0}\left(z \Big|\begin{array}{c} a \\ b, c \end{array}\right) = z^b e^{-z} U(a - c, b - c + 1, z)$ [25] (where $U(.,.,.)$ is the confluent hypergeometric function defined in [33, eq. (13.2.5)]), (8) can be written as

$$f_Y(y) = \frac{\sqrt{\pi}\beta^{n\alpha}}{\Gamma^2(n\alpha)}\left(\frac{y}{2}\right)^{n\alpha-1}e^{-2\beta y}U\left(\frac{1}{2}-n\alpha, 1-n\alpha; 2\beta y\right) \tag{11}$$

Also, with the help of the fact that [25, eq. (7.621.6)]

$$\int_0^\infty t^{b-1}U(a, c; t)e^{-st}dt$$
$$= \frac{\Gamma(b)\Gamma(b-c+1)}{\Gamma(a+b-c+1)}s^{-b}$$
$$\times {}_2F_1\left(a, b; a+b-c+1; 1-s^{-1}\right) \tag{12}$$

where ${}_2F_1(.;.;.)$ is Gauss hypergeometric function defined in [33, eq. (15.1.1)], the $n$-th moment of $Y$ can be evaluated as

$$\mathbf{E}\left(Y^n\right) = \frac{\sqrt{\pi}\beta^{n\alpha-1}\Gamma(n\alpha + n)\Gamma(2n\alpha + n)}{2^{n\alpha}\Gamma^2(n\alpha)\Gamma\left(n\alpha + n + \frac{1}{2}\right)} \tag{13}$$

Notice that the Gauss hypergeometric function of (12) is equal to 1 when the last argument is equal to zero. Let's now use the transformation of variables of $f_Z(z) = 2\,f_Y(2\,z)$ and $F_Z(z) = F_Y(2\,z)$, where the RV $Z = Y/2$, i.e., $Z = \gamma^{1/n}$. since the CDF of the instantaneous end-to-end SNR through the $i$-th path, $\gamma_i$ is a continuous monotonically increasing function, from (9) and (11), with $\alpha = m/n$, and $\beta = 2\,m/\left(\Omega\bar{\gamma}^{1/n}\right)$,

the PDF and the CDF of the $i$-th instantaneous SNR $\gamma_i$ can be found with the help of [31, Sec. 5.1, Sec. 5.2], as

$$f_{\gamma_i}(\gamma) = \frac{2\sqrt{\pi}\,\beta_i^{m_i}}{n_i\,\Gamma^2(m_i)} \gamma^{\frac{m_i}{n_i}} e^{-4\beta_i\gamma^{\frac{1}{n}}}$$
$$\times U\left(\frac{1}{2} - m_i, 1 - m_i, 4\beta_i\gamma^{\frac{1}{n_i}}\right) \quad (14)$$

and

$$F_{\gamma_i}(\gamma) = \frac{\sqrt{\pi}\,\beta_i\,\gamma^{\frac{1}{n_i}}}{2^{2\,m_i-3}\Gamma^2(m_i)} G_{2,3}^{2,1}\left(4\beta_i\gamma^{\frac{1}{n_i}} \Big|\, \begin{matrix} 0, m_i - \frac{1}{2} \\ m_i - 1, 2m_i - 1, -1 \end{matrix}\right) \quad (15)$$

Using (13) with the fact $\Gamma(2\alpha) = 2^{2\alpha-\frac{1}{2}}\Gamma(\alpha)\Gamma\left(\alpha + \frac{1}{2}\right)$ [32 eq. (6.1.18)], and [31, Sec. 5.3], we obtain the approximate average SNR $\bar{\gamma}_i$ as

$$\bar{\gamma}_i = \frac{2^{2m_i-1}(m_i/\Omega_i)^{m_i-1}(m_i)_{n_i}(2m_i)_{n_i}}{(m_i + \frac{1}{2})_{n_i}}\,\bar{\gamma}^{\frac{1-m_i}{n_i}} \quad (16)$$

where $(x)_n = \Gamma(x+n)/\Gamma(x)$. Replacing ( 15) in (2), we can calculate the CDF lower-bound of the $i$-th order statistics $\gamma_{(i)}$ As a special case of the general result for (2), the CDF of the largest order statistics of a random sample is also determined as $F_{\gamma_{(N)}}(\gamma) = \prod_{i=1}^{N}[\gamma_i]$.

## V. PERFORMANCE ANALYSIS

### A. Outage Probability

The outage probability of channel is defined as the probability that the received SNR $(\gamma_{(i)})$ at the destination falls below a predetermined threshold value, namely $P_{out} = \Pr(\gamma_{(i)} \leq \gamma_o)$.

*1) S-DF Relaying:* The outage probability $P_{out} \triangleq F_{(i)}(\gamma_o)$ for the S-DF relaying scheme is expressed as (2). On the other hand, the outage probability can be defined as the probability that the maximum SNR, $\gamma_{(N)}$, at the destination node falls below the threshold value, as (6)

$$P_{out} = F_{\gamma_{(N)}}(\gamma_o) \quad (17)$$

Fig. 3 shows $P_{out}$ for the S-DF relaying scheme over cascaded Rayleigh fading channels ($n = 2, 3, 4, 5$) at a fixed number of relays (e.g., $N = 5$). From Fig.3, there is an excellent match between the approximate and exact results (e.g., based on (4) and (1)). So, if the double Rayleigh acts as a reference point, we observe that the outage probability degrades for larger cascading order $n$. Specifically, at $P_{out} = 10^{-3}$, a performance loss of 5.6, 10.6, 15dB is observed for $n = 3, 4$, and 5 respectively. Based on these observations, it is important to take into account the dynamic range of measurement devices for detecting symbols over such severe fading channels. For instance, when the outage probability based on (6) is assumed at $P_{out} = 10^{-3}$, the required minimum SNR $(P/N_o)$ levels for receiving an undistorted signal are 22.6, 28.2, 33.2, and 37.6 dB for $n = 2, 3, 4$, and 5, respectively. A radio receiver with limited dynamic range will lead to amplitude distortion.

*Asymptotic Analysis:* In order to gain further insight into the performance over $n$*Rayleigh fading channels, we present an asymptotic analysis for $P_{out}$ over i.i.d random variables (i.e., $F_{\gamma_{i1}}(\gamma_o) = F_{\gamma_{i2}}(\gamma_o)$), which provides the maximum achievable diversity order ($d$) of the underlying scheme. By (3) we have $F_{\gamma_i}(\gamma_o) \leq 2\,F_{\gamma_{i2}}(\gamma_o)$, and (17) is upper-bounded by

$$P_{out} \leq \left(2\frac{\gamma\left(m, \beta\gamma_o^{\frac{1}{n}}\right)}{\Gamma(m)}\right)^N \quad (18)$$

where $\gamma(.,.)$ is the lower incomplete gamma function defined in [25]. Then, at high SNR (i.e., $\bar{\gamma} \to \infty$), with the help of the facts: $\gamma(\alpha, x) = \frac{x^\alpha}{\alpha}M(\alpha, \alpha+1, -x)$ and $M(a, b, x) = 1$ as $|x| \to 0$, given by [33], where $M(.,.,.)$ is the Kummer's confluent hypergeometric function, ( 18) can be written as

$$P_{out} \leq \left(\frac{2^{m+1}(m/\Omega)^m}{m\,\lambda^{m/n}\Gamma(m)}\right)^N \left(\frac{\gamma_o}{SNR}\right)^{\frac{mN}{n}} + O\left(\left(\frac{\gamma_o}{SNR}\right)^{\frac{mN+1}{n}+1}\right) \quad (19)$$

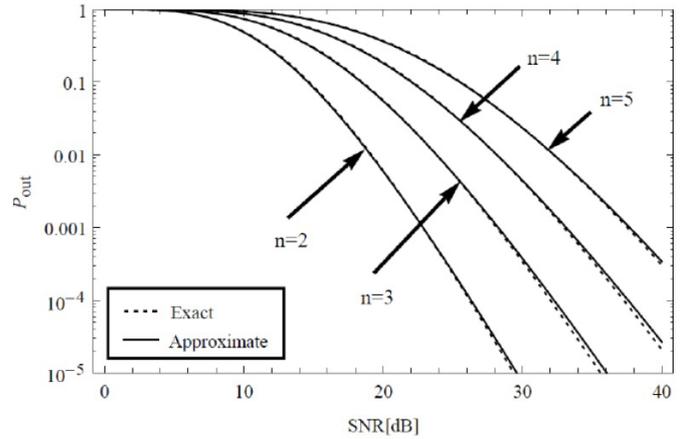

Fig. 3. Outage performance of the S-DF relaying scheme over $n$*Rayleigh fading channels ($N = 5$).

From (19), we can deduce that the maximum diversity order for the S-DF scheme over $n$*Rayleigh fading channels is $d = mN/n$. This is because diversity order is defined as the slope of the $P_{out}$ curve as a function of the average SNR in log-log scale, i.e., [34]

$$d = \lim_{SNR \to \infty}(-\log P_{out}/\log SNR) = \frac{mN}{n} \quad (20)$$

As we note, the result in (20) is novel and generalizes known results on the diversity order of these relaying schemes on Rayleigh fading channels to the cascaded Rayleigh fading scenario. Our analytical results show that the full diversity order ($d \approx N$) can be obtained for classical Rayleigh fading channels (where the selected relay is fixed). In addition, the diversity order is inversely proportional to the cascading order $n$ and improves as $N$ increases. Fig.4. depicts the diversity order over Rayleigh and cascaded Rayleigh fading channels, assuming $N = 2$ and 4. As can be observed, the full diversity order for the Rayleigh fading channel approaches $N$ as SNR tends to infinity, while the diversity order for the cascaded Rayleigh fading channels decreases linearly with increasing $n$ to reach an asymptotic value equivalent to $mN/n$, confirming our analytical results. Furthermore, it is noticed that the increase in the number of relays leads to improved system performance for both Rayleigh and cascaded Rayleigh fading channels, in which the diversity order can be maximized, resulting in lower outage probability. Therefore, cooperative diversity systems can allow accurate symbol detection even using measurement devices with a low dynamic range. From (19), we can also deduce that the effective coding gain (CG) is given by

$$\text{CG} = \left( \frac{2^{m+1}(m/\Omega)^m}{m\lambda^{m/n}\Gamma(m)} \right)^{-\frac{n}{m}} \quad (21)$$

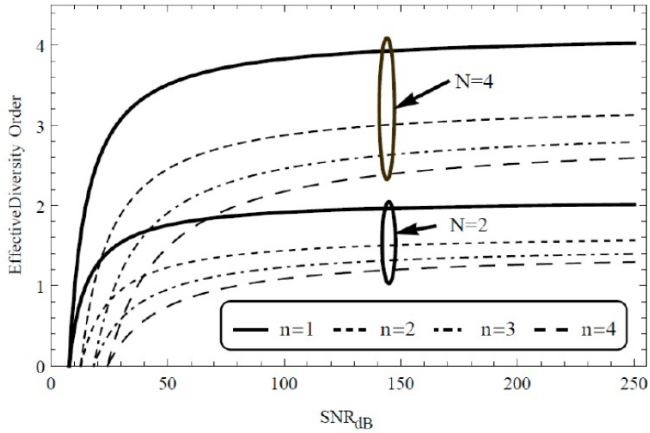

Fig. 4. Effective diversity order for the S-DF relying scheme over Rayleigh and $n$*Rayleigh fading channels.

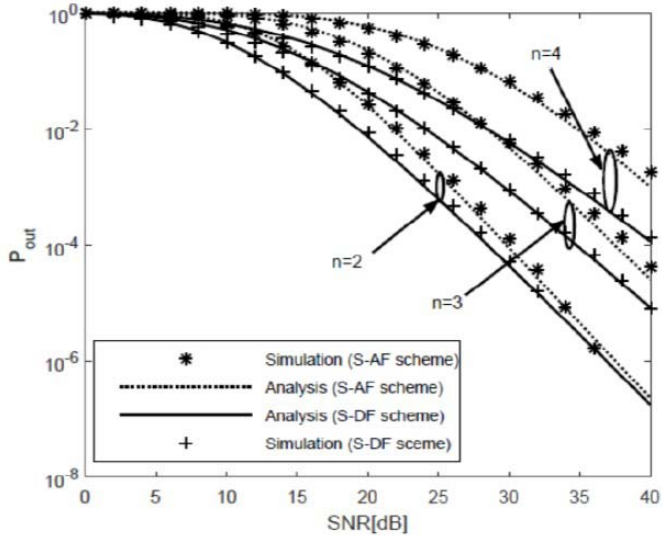

Fig. 5. Comparison of outage probability of the S-AF and the S-DF relaying schemes over $n$*Rayleigh fading channels ($N = 3$).

Note that the coding gain in ( 21) depends only on the fading severity parameters and channel variance which are assumed to be fixed during the entire transmission time, regardless the number of relays $N$.

*2) S-AF Relaying:* In case of the S-AF relaying scheme, the approximate outage probability based on (15) is

$$P_{out} = \prod_{i=1}^{N} F_{\gamma_i}(\gamma_o) \quad (22)$$

In addition, at high SNR levels, we can apply (14) instead of (15) to simplify the analysis of the maximum diversity order achievable over $n$*Rayleigh fading channels, using the facts that [33, eq. (13.5.12)]

$$U(a, b, x) = \frac{\Gamma(1-b)}{\Gamma(1+a-b)} + \mathcal{O}(|x|), \quad |x| \to 0$$

and $\int_0^u x^{v-1}\exp(-\mu x)dx = \mu^{-v}\gamma(v, \mu u)$[25, eq.(3.381.1)], this leads us to rewrite $P_{out}$ (where the random variables are i.i.d) in an asymptotic form as

$$P_{out} = \left[ \frac{1}{2^{2m-1}} \left( \frac{\gamma\left(m, 4\beta\gamma_o^{\frac{1}{n}}\right)}{\Gamma(m)} \right) \right]^N \quad (23)$$

Also, eq. ( 23) can have another asymptotic expansion based on the facts [33, eq. (6.5.12) and (13.5.5)], as

$$P_{out} = \left( \frac{2^{5}\,^{m-1}(m/\Omega)^m}{m\lambda^{m/n}\Gamma(m)} \right)^N \left( \frac{\gamma_o}{\text{SNR}} \right)^{\frac{mN}{n}} + O\left( \left( \frac{\gamma_o}{\text{SNR}} \right)^{\frac{mN}{n}+1} \right) \quad (24)$$

From (24), we can extract the maximum achievable diversity order for the $n$*Rayleigh distribution as $d = mN/n$. As it is expected that both S-AF and S-DF relaying schemes have the same diversity order over an $n$*Rayleigh fading channel.

Fig. 5 compares the outage performance of the S-DF and SAF relaying schemes over $n$*Rayleigh fading channels (i.e., (17) versus (22)). Since the exact expression for the outage probability for the S-AF relaying scheme is very challenging to calculate because of (1), we used the Mont-Carlo simulation against the approximate expression (22). From Fig. 5, there are key points could be extracted as follows:

- There is an excellent agreement between the analytical results and Monte-Carlo simulation. This can clearly be noticed for the S-DF relaying scheme.
- In comparison with the S-AF relaying scheme, the S-DF relaying improves the outage performance at low and high average SNR.
- The tightness between the two schemes is gradually improved for small $n$ values in the high-SNR regime, confirming our earlier observations that the maximum diversity order ($d$) is the same in both schemes at high SNR values.

The major findings summarized above are important when we need to estimate cascaded Rayleigh fading channels associated with the S-AF relaying schemes (where high noise generated around the relays, resulting in harsh keyhole channels between the source and destination). Therefore, choosing the best relay (with high SNR and a low cascading order, e.g., $n = 2$ ) which provides low outage performance is challenging, especially when thinking of a high-mobility vehicle, where the impulse response of an $n$*Rayleigh channel changes rapidly during the symbol period (i.e., a fast-fading channel and a severe drop in SNR), resulting in the target localization problem.

### B. Power Allocation Optimization

In the context of inter-vehicular communication, optimizing the power among the source and the relays is critical to reduce the total transmit energy. In practice, the cascaded Rayleigh channel coefficients ($h_{i1}, h_{i2}$) can be estimated and then used to detect the signal at the destination. Relay nodes that operate in the DF mode also require channel knowledge in the source relay link $s \rightarrow r_i$ to decode the source signal. For AF relaying, the knowledge of $n$*Rayleigh fading channels at the relay nodes is required for appropriately scaling the received signal to satisfy relay power constraints; therefore, the quality of the channel estimation process can generally affect the overall performance of cooperative transmission and may become a performance limiting factor for IVC systems. In general, the fading channel coefficients can be acquired by either blind techniques or through the use of pilot symbols methods [35] In practical terms, blind channel estimation techniques suffer from several disadvantages, such as the high computational complexity and slow convergence, which are prohibitive for high-mobility vehicle scenarios. However, due to the complexity of power allocation (PA) for the S-AF relaying schemes as noted in (22), we analyze the optimal PA for the case of S-DF when only statistical CSI, ($\lambda_{i1}, \lambda_{i2}$), is available at the source and relays rather than instantaneous CSI. By doing so, we reduce the outage probability under the total power constraint of $P_1 + P_2 \leq P_T$ where $P_1$ is the radio transmit power of the source signal, $P_2$ is the transmit power of the selected relay, and $P_T$ is the total transmit power. Here, in this scenario, we assume that all relays use the same transmit power $P_2$; hence, based on (6) the optimization problem can be formulated as follows

$$\min_{P_1, P_2} \sum_{i=1}^{N} \left[ 1 - \frac{\Gamma\left(m_{i1}, \beta_{i1}\gamma_0^{\frac{1}{n_{i1}}}\right) \Gamma\left(m_{i2}, \beta_{i2}\gamma_0^{\frac{1}{n_{i2}}}\right)}{\Gamma(m_{i1}) \Gamma(m_{i2})} \right]$$

$$\text{s.t. } P_1 + P_2 \leq P_T \text{ and } P_1, P_2 \geq 0 \quad (25)$$

where $\bar{\gamma}_{i1} = \lambda_{i1}P_1/N_o$, $\bar{\gamma}_{i2} = \lambda_{i2}P_2/N_o$

By applying the method of Lagrange multipliers, the PA for the source is derived as

$$P_1 = \sum_i \frac{P_{\text{out}}\mu_i \beta_{i1}^{m_{ij}} \gamma_o^{\frac{1}{n_{i1}}} e^{-\beta_{i1}\gamma_0^{\frac{1}{n_i}}}}{n_{i1}\Gamma(m_{i1})(1 - \mu_i \omega_i)\,\xi} \quad (26)$$

where $\omega_i = \frac{\Gamma\left(m_{i2}, \beta_{i2}\gamma_o^{\frac{n_{i2}}{2}}\right)}{\Gamma(m_{i1})}$, $\mu_i = \frac{\Gamma\left(m_{i2}, \beta_{i2}\gamma_o^{\frac{n_{i2}}{2}}\right)}{\Gamma(m_{i2})}$

A similar equation can be derived for the selected relay, $P_2$ Using (26) and setting $P_2 = P_T - P_1$, the approximate power allocation for $P_1$ can be written in the following form

$$P_1 = P_T \left[ \frac{\sum_i \omega_i \theta_{i2}\beta_{i2}^{m_{i2}} \gamma_o^{\frac{m_{i2}}{n_{i2}}} e^{-\beta_{i2}\gamma_o^{\frac{1}{n_{i2}}}}}{\sum_i \mu_i \theta_{i1}\beta_{i1}^{m_{i1}} \gamma_o^{\frac{m_{i1}}{n_{i1}}} e^{-\beta_{i1}\gamma_o^{\frac{1}{n_{i1}}}}} + 1 \right]^{-1} \quad (27)$$

where $\theta_{ij} = 1/n_{ij}\Gamma(m_{ij})(1 - \omega_i \mu_i)$. It should be noted that (27) is a transcendental function and it is challenging to find a closedform for the source power. Thus, we calculate it numerically using a root-finding algorithm such as Bisection, Newton or successive numerical approximation methods.

At this stage, given the total power constraint, the source and the selected relay power can be set to an output $P_1 = \rho P_T$ and $P_2 = (1 - \rho)P_T$, where $\rho$ is the PA ratio ($\rho \in (0, 1)$), which is calculated from (27) through the successive approximation algorithms [36, section 14.1]

*Asymptotic Solution:* A simple asymptotic solution for (27) can be determined by using the fact that $x^\alpha e^{-x} = \Gamma(\alpha+1, x) - \alpha\Gamma(\alpha, x)$[25], and by noting that $\Gamma(\alpha+1, x) \leq \alpha\Gamma(\alpha)$, where $x$ is sufficiently small. In this case, the optimization problem can be rewritten in a simple compact form as

$$P_1 = P_T \left[ \frac{\sum_i m_{i2}/n_{i2}}{\sum_i m_{i1}/n_{i1}} + 1 \right]^{-1} \quad (28)$$

From (28), it can be seen that the PA for the source depends on the fading severity parameters, regardless of the channel statistics ($\lambda_{i1}, \lambda_{i2}$). Physically speaking, when the source and the best relay's power is high, the effect of path loss (i.e., $\lambda_{i} \propto d_{ij}^{-\sigma}$, where $d_{i1}$ represents the distance between the source and the relay, $d_{i2}$ is the distance between the relay and destination, and $\sigma$ is the path loss exponent) is negligible, corresponding to the same scenario when the selected relay is located in the middle between the source and the destination, resulting in a similar path loss on both terminals.

*1) Machine Learning-Based PA:* In order to classify the fading severity parameter $n$ in IVC networks, we use a simple machine learning algorithm such as Naive Bayes; it's a simple probabilistic classifier that requires only a small number of training data $K$ to estimate the required parameters for classification. Therefore, for each of $n$ possible classes $C_{ij}^n$, we need to calculate the conditional probability of $\Pr\left(C_{ij}^n|h_{ij,1}, \ldots, h_{ij,K}\right) \propto \Pr\left(C_{ij}^n\right)\prod_{t=1}^{K}\Pr\left(h_{ij,t}|C_{ij}^n\right)$, where we make the assumption that $h_{ij,1}$ through $h_{ij,K}$ are conditionally independent given a class label $C_{ij}^n$. Now using the maximum a posteriori (MAP) estimation, the Bayes classifier assigns a class label $C_{ij}^n$ for each $n$*Rayleigh fading channel $h_{ij}$, as

$$\hat{C}_{ij}^n = \underset{n_{ij}, \sigma_{ij}^2}{\operatorname{argmax}} \Pr\left(C_{ij}^n\right)\prod_{t=1}^{K}\Pr\left(h_{ij,t}|C_{ij}^n\right) \quad (29)$$

where $\Pr\left(C_{ij}^n\right)$ is the prior probability of the class variable $C_{ij}^n$ which can be identified through real-time data measurements for $n$*Rayleigh channels [11]. The probability density of the $n$*Rayleigh distribution given a class $C_{ij}^n$, is computed by

$$\Pr\left(h_{ij} = h|C_{ij}^n\right) = 2\left(\frac{m_{ij}}{\Omega_{ij}}\right)^{m_{ij}} \frac{h^{2\alpha_{ij}-1}}{n_{ij}\Gamma(m_{ij})\sigma_{ij}^{2\alpha_{ij}}} e^{-\beta_{ij}h^{\frac{2}{n_{ij}}}}$$

$$(30)$$

TABLE I

A Set of Labelled Training Data Is Used to Estimate the n*Rayleigh Distribution, Where ffy $= 2^{-n}$' A Monte-Carlo Simulation Is Performed at $K = 10^6$ Samples. The Table Shows Some Estimated Samples of the Empirical PDFs

| Class Label, $C_{ij}^n$ | | $h_{ij,1} = 0.12$ | $h_{ij,2} = 0.24$ | $h_{ij,3} = 0.54$ | $h_{ij,4} = 0.78$ | $h_{ij,5} = 1.0$ | $h_{ij,6} = 1.23$ |
|---|---|---|---|---|---|---|---|
| $n_{ij}$ | $\sigma_{ij}^2$ | | | | | | |
| 1 | 0.50 | 0.238 | 0.449 | 0.813 | 0.855 | 0.746 | 0.538 |
| 2 | 0.25 | 0.756 | 0.923 | 0.813 | 0.619 | 0.478 | 0.319 |
| 3 | 0.125 | 1.289 | 1.167 | 0.703 | 0.467 | 0.238 | 0.229 |
| 4 | 0.063 | 1.662 | 1.167 | 0.595 | 0.375 | 0.249 | 0.212 |
| 5 | 0.031 | 1.883 | 1.167 | 0.507 | 0.298 | 0.201 | 0.135 |
| 6 | 0.016 | 1.957 | 1.092 | 0.424 | 0.241 | 0.159 | 0.112 |

where $\alpha_{ij} = \frac{m_{ij}}{n_{ij}}$, $\beta_{ij} = \frac{m_{ij}}{\Omega_{ij}} \sigma_{ij}^{-2/n_{ij}}$. The maximum likelihood (ML) estimator[2] for the parameter $\sigma_{ij}^2$ is given by

$$\hat{\sigma}_{ij}^2 = \left[ \frac{\sum_{t=1}^K h_t^{2/n_{ij}}}{2K\,\Gamma^{n_{ij}}\left(1/n_{ij} + 1\right)} \right]^{n_{ij}} \qquad (31)$$

Therefore, if we suppose that the training fading data listed in Table I (where $K$ experimental samples need to be collected from different environments (e.g., large cities, small cities, and highways [11], [16], [37] contains a continuous attribute $h$, then the probability distribution of $h$ given a class $C_{ij}^n$, is calculated by (30), which in turn makes the Bayes classifier of (29) to assign a class label for the variable $h$ (where we can localize the best relay associated with the lowest cascading order $n$ e.g., $C_{ij}^n$ for $n = 1, 2$) and perform an efficient PA scheme for the source- $P_1$ and selected relay-$P_2$ that can reduce the outage probability. If the IVC network is operated at the sub-optimality (i.e., the conditional independence assumption is violated), the zero-one loss function [38] does not penalize inaccurate probability estimation as long as the maximum probability is set to the correct class. This means that Naive Bayes may change the posterior probabilities of each n*Rayleigh fading channel class, but the class with the maximum posterior probability is often unchanged. Thus, the classification is still correct, although the probability estimation is poor [39].

In Fig.6, we evaluate the impact of the S-DF relay systems (PA-based) on the cascaded Rayleigh fading channels. In this example, two main modes of transmission are compared: the PA mode under statistical CSI, and the equal power allocation (EPA) where the total transmitted power $P_T$ is divided equally between the source and the selected relay (i.e, $P_1 = P_2 = P_T/2$). We assume that the channel quality between the selected relay and the destination is much better than that between the source and the selected relay (e.g., $\lambda_1 = 1$ and $\lambda_2 = 10$). As observed from the figure, the PA mode has an advantage over the EPA mode by reducing the outage probability. This is mainly because the PA mode devotes larger power to the weaker link to reduce the overall outage probability. In this case, the PA ratio $\rho = P_1/P_T$ is evaluated from (27) using the successive approximation algorithm, which yields $\rho = 0.757, 0.629, 0.534$ and $0.462$ for $n = 1, 2, 3$ and 4 respectively. It should be noted here that the PA ratio converges to 0.5 when $n$ increases, which means that the EPA policy is near-optimal for $n \geq 3$. In this case, we can use EPA instead of PA to get a lower outage probability. Since the distance between the transmitter and the receiver is another important factor for keyhole channels, we investigate the EPA mode in terms of distance. Hence, we can redefine the instantaneous SNR as $\bar{\gamma}_{i1} = d_{i1}^{-\sigma} P_1/N_o$, $\bar{\gamma}_{i2} = d_{i2}^{-\sigma} P_2/N_o$. Fig.7 shows the outage probability (based on (6)) versus distance between the source and the selected relay, $d_1 = 1 - d_2$, with SNR $= 20$dB and a path loss exponent of $\sigma = 3$ (this could describe, for example, an urban scenario). As is obvious from the figure, the outage probability increases when the distance between the source and the selected relay $d_1$ increases for both Rayleigh and n*Rayleigh fading channels, but the degree of concavity of the curves is gradually decreased by increasing the value of $n$. Furthermore, Fig.7 shows that the minimum outage probability occurs at distance $d_1 = 0.5$. This is indeed expected because the system performance is limited to the channel quality over the source to relay links.

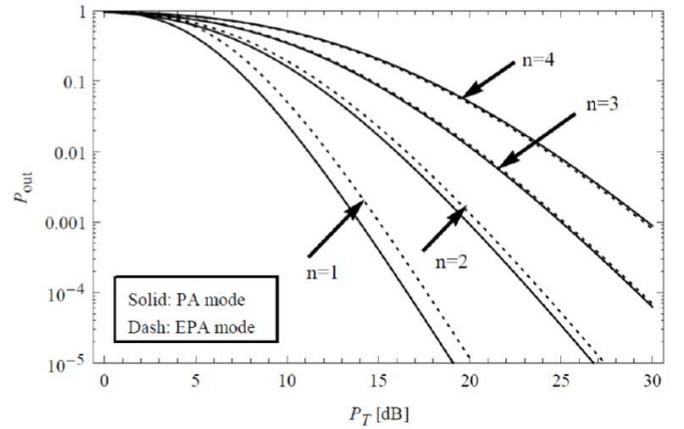

Fig. 6. Effect of the PA and EPA modes on the outage performance of the S-DF relaying scheme over Rayleigh and n*Rayleigh fading channels.

---

[2] We use the maximum likelihood method in finding the parameters $\sigma_{ij}^2$ that maximize the likelihood of the observed data set, $h_{ij}$, and make the Naive Bayes model fits the n*Rayleigh distribution (30).

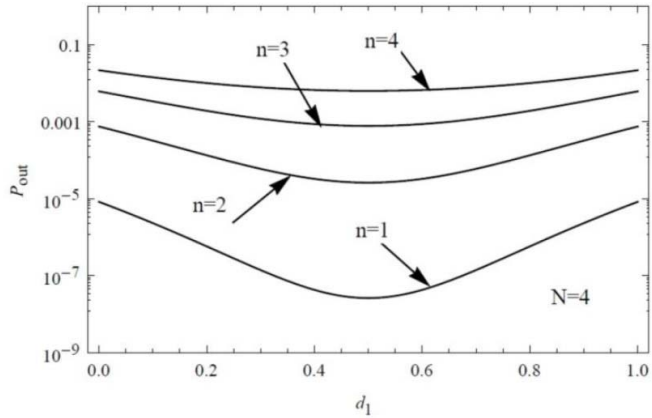

Fig. 7. Outage performance of the S-DF-relaying scheme over Rayleigh and $n$*Rayleigh fading channels in term of the distance between the source and the selected relay.

TABLE II
SIMULATION PARAMETERS

| Parameter | Value |
|---|---|
| Data rate | 2Mbps (UDP) |
| Packet generation rate | $10 - 60$ packets/second |
| Packet size | 512 byte |
| Transmission power | 20 mW |
| Sensitivity | -89 dBm |
| Duration of a time slot | 13 $\mu s$ |
| Transmission range | 250 m |
| Beacon rate | 1 Hz |
| Speeds | 30 and 80 km/h |

## VI. EXPERIMENTAL PERFORMANCE EVALUATION

In this section, we provide experimental and simulation results to show the optimal relay selection time (convergence time) for both S-AF and S-DF schemes. We test the packet delivery ratio and outage probability for the S-DF schemes across the PA and EPA modes, and use the network simulator 2 (NS-2) and simulation of urban mobility (SUMO) to simulate vehicular ad hoc networks (VANET). To make the evaluation realistic, we run a simulation using a map of Montreal which has a grid road topology with urban highways and bidirectional roads (e.g., map of the $3.16km \times 3.16km$ area; see Fig.8 and simulation parameters given in Table II). We set the speed of vehicles between 30 and 80 km/h, which is common for the city environment. The road topology is obtained using OpenStreetMap and is filtered, formatted, and converted into a SUMO network file. Using SUMO, vehicular mobility traces are generated and used to populate the chosen simulated area (e.g., urban scenarios), where local scattering objects (e.g., buildings, vehicles, road signs, bridges, street corners, trees, bridges, hallways, etc) are considered. In our simulation, we chose a moderately sparse network to have a moderate simulation processing time (e.g., we simulate 600 vehicles per $10km^2$, i.e., the vehicle density is 60 veh/km² with a radio range of 250m). Each vehicle is equipped with a dedicated short-range communication (DSRC) transceiver to enable vehicle-to-vehicle (V2V) and vehicle-to-infrastructure (V2I) communications, where some road side units (RSUs) (30 RSUs) are deployed within the simulation area following a uniform distribution. In order to assess the efficiency of the selected relays, we randomly choose the source and destination from a set of vehicles where the destination (e.g., vehicle, RSU) is within one or two hops from the source. Here, the source is sending packets via user datagram protocols (UDP) protocols. If the source wants to reach the destination through a relay because of the weak direct communication (i.e. low SNR), our algorithm will select the optimal relay that has the best link quality (i.e., high SNR) with the destination, regardless of the relay position, whether it is at an intersection or a turn. As an example of application scenario, this can be used as a solution for cooperative vehicle and infrastructure system (CVIS) to enable better V2V and V2I communications. The proposed mobile relay solution will extend RSU and LTE-V small cells coverage and enhance data transmission. For CVIS applications, this can assist data aggregation for mobile edge computing which in turn improves vehicle positioning and local dynamic maps.

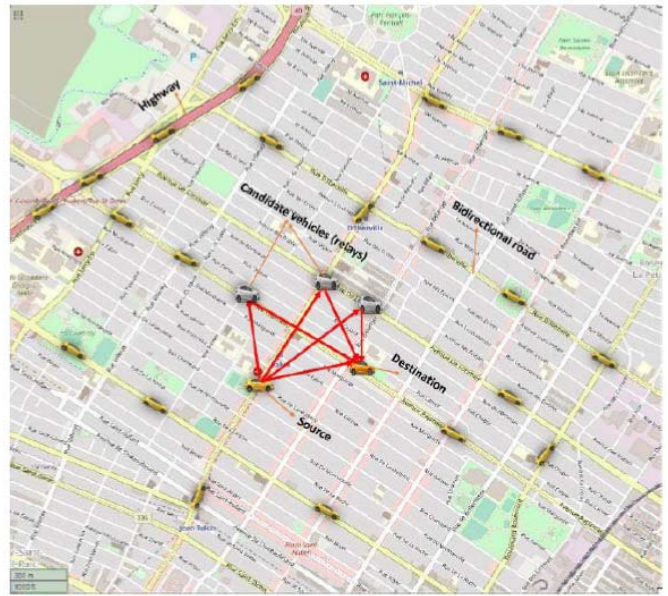

Fig. 8. Traffic scenario in Montreal that includes a grid road topology (e.g., map of the 3.16km×3.16km area) with urban highways and bidirectional roads, where the source sends his message to a set of relays (i.e., candidate vehicles) and the destination (e.g., vehicle, RSU). Here, the S-DF relaying policy is applied to choose the most reliable path $s \rightarrow r_i \rightarrow d$ over $n$*Rayleigh fading channels and forward the source message to the destination.

In this scenario, we implement both relay selection strategies; the S-DF and S-AF schemes over the $n$*Rayleigh distribution. Note that 600 vehicles are the number of vehicles simulated in the selected urban area and not the candidate vehicles (relays) used in our relay selection algorithms. In this setting, we generally seek to evaluate two things: 1) the performance of S-DF schemes over different metrics (outage probability and packet delivery ratio) and compare the simulation results with the analytical results. 2) examine the processing time needed to select the optimal relay based on the training data of machine learning-based PA algorithms (where Monte-Carlo simulation is performed at $K = 10^{10}$ samples). In Fig.9, we evaluate the S-DF scheme over the PA and EPA modes in terms of packet delivery ratio and $P_{out}$. In Fig.9 (a) and (b), we observe that PA outperforms EPA in terms of packet delivery ratio over different distances and packet generation rates via $n$*Rayleigh fading channels (e.g., $n = 1, 2, 3, 4$), where not only the optimal channel is selected but also the optimal relay of each selected channel, even if the distance between the source and the destination increases. Also, we can show that the PA algorithm gives a stable delivery ratio of packets around 85% and 95% compared with the EPA mode where the ratio is around 50% only. In Fig.9 (c), we test the packet delivery ratio under various vehicle densities, where the PA performance gradually approaches the EPA when the vehicle density is high; this is due to the presence of NLOS propagation paths and increasing the cascading order ($n$) between the selected relay and the destination (i.e., poor channel quality and low SNR) with an increase in vehicle density, which leads to a degradation of the PA performance. In Fig.9 (d), we demonstrate the outage probability of the the relay selection algorithm compared to direct transmission scenarios over $n$*Rayleigh fading channels, where we set the vehicle density at 60 veh/km². The results show that the proposed algorithm can offer large power or energy savings compared to direct transmission, where the algorithm can optimize the transmission power between the source and the selected relay and reduce the outage probability in dual-hop IVC systems [40]. Also, we observe that when the distance between the source and the destination increases, the outage probability increases, and when the number of relays increases, $P_{out}$ is reduced. This finding confirms our analytic results of the effect of PA on the outage probability of S-DF schemes.

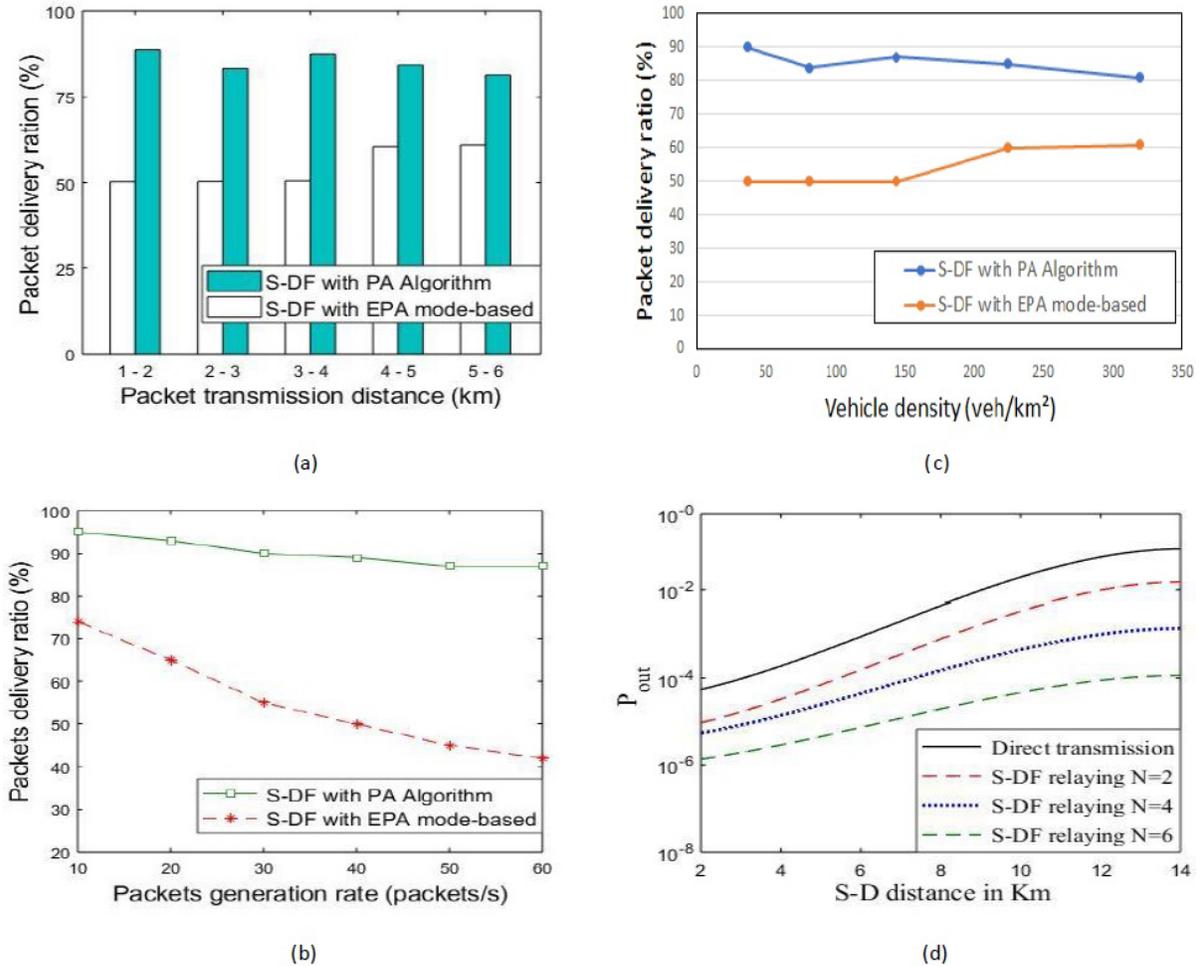

Fig. 9. Evaluation of the S-DF scheme in terms of packet delivery ratio and $P_{out}$ (where both PA and EPA modes are implemented over $n$*Rayleigh fading channels (e.g., $n = 1, 2, 3, 4$ )).

In Fig.10 (a), we provide the processing time (convergence time) for the relay selection algorithm under different vehicle density scenarios, where we set the number of candidate vehicles (relays) $N$ to 4 and Monte-Carlo simulation is performed at $K = 10^4$ samples. Here, we can define the convergence time as the time to select the optimal relay between the candidate nodes that are in the radio range of the source node. Although the relay selection algorithm works well at low cascading orders $n \leq 3$, it should be noted that there are some limitations associated with the algorithm at the cascading order of $n = 4$. By increasing the cascading order $n$ the convergence time becomes larger as the number of vehicles (between the source and destination) increases and the search for the optimal relay is longer than the case of $n = 1, 2,$ and 3, which requires designing an advanced relay selection algorithm that can overcome these constraints and improve the performance of the proposed algorithm. To take a closer look at the performance of the S-DF scheme (PA-based), we compare with other relay selection schemes (such as S-AF) as shown in Fig.10 (b). We observe that the S-DF scheme outperforms the S-AF scheme in terms of convergence time to select the best relay through *n*\*Rayleigh fading channels, where the S-DF scheme can reduce end-to-end transmission delay by selecting the best channel quality between the relay and destination, which has the highest SNR and lowest convergence time. The end-to-end delay can be defined as the convergence time generated by the relay selection algorithm plus the standard transmission and processing time of DF schemes [41], [42]. It is worth mentioning that our application is not a delay-tolerant application like a carry-and-forward strategy [43], where the vehicle speed and transmission direction are important for performance evaluation, as the delay may last for several seconds. In our application, the delay in transmitting and processing the packet is in the order of milliseconds.

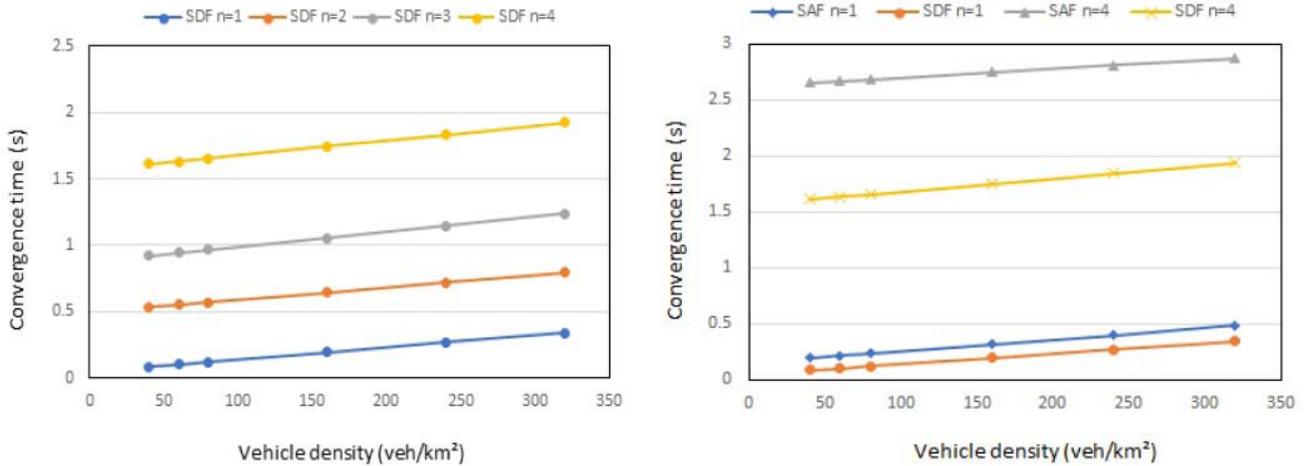

Fig. 10.   (a) Convergence time of the S-DF scheme over *n*\*Rayleigh fading channels. (b) Comparison of convergence time of the S-AF and the S-DF relaying schemes over *n*\*Rayleigh fading channels (where simulation is run under different vehicle density scenarios with a certain number of candidate vehicles (relays), $N = 4$).

Given that the simulation was performed on a computer with a Corei7 CPU and 8GB RAM, and that this type of wireless physical layer algorithms is typically performed on an ASIC baseband processor, the convergence time can be reduced by a factor of 10 to 100 compared to general purpose processors (CPUs) [44]. Hence, based on our results shown in Fig.10 and Table III, the convergence time can be reduced by at least a factor of 10 using ASIC, so that the processing time for 60 vehicles is less than 11ms for $n = 1, 56$ms for $n = 2, 95$ms for $n = 3$, and 164ms for $n = 4$. In order to reduce the convergence time further, we urge to use graphic processing units (GPUs). Over the past two decades, GPUs have become increasingly faster and cheaper. GPUs are progressively incorporated in emerging smart vehicles for assisted and autonomous driving. They are considered as the brain of on-board units for AI algorithms and complex sensor-based systems (such as RADAR, LIDAR, video processing). Moreover, the GPU has become a key component of software-defined radio (SDR), which enables the creation of a flexible and cost-effective V2X communication platform [45]. Compared to our CPU that provide only about 5 gigaflops, basic GPU boards like "NVIDIA Jetson Nano" (which costs just \$99/unit) can deliver up to 472 gigaflops (i.e., \$0.2/gigaflop), which in turn translates to an increase of processing speed of about 100 times. Furthermore, we can get a solution (e.g. Radeon RX 570) that offers 5000 gigaflops for \$130/unit (i.e., \$0.026/gigaflop), which translates to an increase in processing speed of about 1000 times (i.e., the convergence time in Fig.10 becomes between 2 to 30ms for 500 gigaflops, and between 0.2 to 3ms for 5000 gigaflops) [45]. With the rapidly evolving capabilities of digital electronics and large-scale adoption of GPUs for vehicular application processing, the cost of gigaflop is expected to decrease significantly in the coming years.

TABLE III
CONVERGENCE TIME FOR BOTH S-AF AND S-DF SCHEMES
(WHERE PROCESSING TIME IS MEASURED IN SECONDS)

| Vehicle density (veh/km²) | S-AF $n = 1$ | S-DF $n = 1$ | S-AF $n = 4$ | S-DF $n = 4$ |
|---|---|---|---|---|
| 40 | 0.1955564 | 0.0822279 | 2.654228 | 1.6113563 |
| 60 | 0.2159872 | 0.1006942 | 2.672944 | 1.6345374 |
| 80 | 0.2364128 | 0.1191558 | 2.685656 | 1.6577126 |
| 160 | 0.3181308 | 0.1930163 | 2.748516 | 1.7504311 |
| 240 | 0.3998436 | 0.2668721 | 2.811372 | 1.8431437 |
| 320 | 0.4815564 | 0.3407279 | 2.874228 | 1.9358563 |

## VII. CONCLUSION

In this work, we presented a comprehensive performance analysis for the selective DF and AF relaying schemes over $n*$Rayleigh fading channels. The outage probability and diversity order have been analyzed for the considered schemes. Our analysis and simulation results have shown that the relay selection technique achieves a maximum diversity order of $mN/n$, which can be considered valuable guidelines for engineers working on the design of measurement devices for cascaded Rayleigh fading channels. The diversity order is the asymptotic slope of the outage probability curve, and thus a high diversity order means that the outage probability decreases faster with increasing SNR. This then means that at a particular (high) SNR, a system with higher diversity order is more reliable. Our paper shows that the S-AF and S-DF approaches both exhibit the highest possible diversity order given the number of relays deployed. In practical terms, this means more reliable links for a given (high) SNR and this therefore translates to an increased level of safety. In general, our statistical analysis showed that machine learning plays a key role in selecting the best relay and allocating energy. The results confirm that transmit power allocation optimization is required for IVC systems when the cascading order $n \leq 2$. Plus, the time required to find the optimal relay is greatly reduced when the cascading order $n$ decreases. Of course, this study will help automakers deploy a dynamic IVC network that can significantly improve safety and operational efficiency.

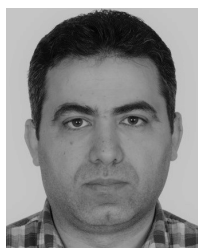

**Yahia Alghorani** (Member, IEEE) received the M.Sc. degree in computer engineering from University Kebangsaan Malaysia, Malaysia, in 2007, and the Ph.D. degree in computer and software engineering from the Ecole Polytechnique de Montreal, Montreal, QC, Canada, in 2015. From 2009 to 2010, he was an RF Engineer with Batelco Telecommunication Company. From 2010 to 2012, he was a Wireless Networks Engineer with Huawei Technologies Company Ltd. From 2017 to 2019, he was a Research Fellow with University College Dublin, Ireland. He is currently a Research Associate with Lakehead University, Thunder Bay, ON, Canada. His research interests include intelligent transportation systems, connected vehicles, artificial intelligence, machine learning, deep learning, medical wearables and e-health, statistical signal processing, and hardware/software co-design.

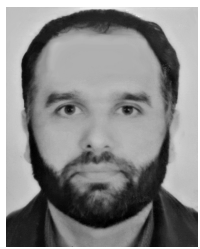

**Ahmed Salim Chekkouri** (Member, IEEE) received the M.Sc. degree in electrical engineering from the University of Sherbrooke, Sherbrooke, QC, Canada, and the Ph.D. degree in computer engineering from the Ecole Polytechnique de Montreal, University of Montreal, Montreal, QC, Canada, in 2003 and 2015, respectively. He was an Adjunct Researcher with the Mobile Computing and Networking Research Laboratory (LARIM) Chair Ericsson Canada. In 2017, he was a Post-Doctoral Fellow with the Department of Computer Engineering, Ecole Polytechnique de Montreal. He is currently a Senior IT Consultant with Medunicom, Ecole Polytechnique de Montreal. His research interests include the Internet of Things and mobile networks, AI and machine-learning, embedded systems and sensors for ITS and e-Health, and cybersecurity.

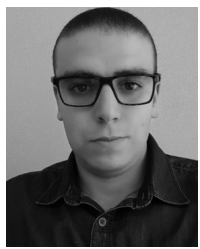

**Djabir Abdeldjalil Chekired** (Student Member, IEEE) received the B.S. degree in computer science and the M.S. degree in computer networks and distributed systems from the University of Science and Technology Houari Boumediene (USTHB), Algiers, Algeria, in 2010 and 2012, respectively. He is currently pursuing the joint Ph.D. degree in computer science with the Environment and Autonomous Networks Laboratory (ERA), University of Technology of Troyes, France, and the School of Electrical Engineering and Computer Science, University of Ottawa, Canada. His research interests include new cloud computing design for smart grid systems and analysis in green electric vehicles networks, smart grid energy management, fog-SDN architectures, and the Internet of Things.

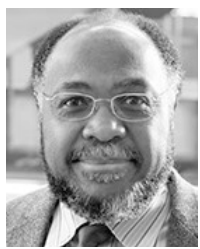

**Samuel Pierre** (Senior Member, IEEE) received the B.Eng. degree in civil engineering from the Ecole Polytechnique de Montreal, Montreal, QC, Canada, in 1981, the B.Sc. and M.Sc. degrees in mathematics and computer science from the Université du Québec à Montréal (UQAM), in 1984 and 1985, respectively, the M.Sc. degree in economics from the University of Montreal in 1987, and the Ph.D. degree in electrical engineering from the Ecole Polytechnique de Montreal in 1991. From 1987 to 1998, he was a Professor with the University of Québec at Trois-Rivières, Trois-Rivières, QC, USA. Prior to joining the Télé-Université of Québec, he was an Adjunct Professor with the Université Laval, Quebec City, QC, Canada, and an Invited Professor with the Swiss Federal Institute of Technology, Lausanne, Switzerland, and Université Paris 7, Paris, France. He is currently with the Ecole Polytechnique de Montreal, where he is also a Professor of computer engineering; the Director of the Mobile Computing and Networking Research Laboratory (LARIM), Department of Computer Engineering; the NSERC/Ericsson Chair of next generation and mobile networking systems; and the Director of the Mobile Computing and Networking Research Group (GRIM). His research interests include wireline and wireless networks, mobile computing, artificial intelligence, and telelearning. Dr. Pierre is a fellow of the Engineering Institute of Canada. He is an Associate Editor of the IEEE COMMUNICATIONS LETTERS, the IEEE CANADIAN JOURNAL OF ELECTRICAL AND COMPUTER ENGINEERING, and the IEEE CANADIAN REVIEW. He is a Regional Editor of the *Journal of Computer Science*. He serves on the Editorial Board of *Telematics and Informatics*, which are edited by Elsevier *Science*.